\documentclass{amsart}

\usepackage[dvips]{graphicx}
\usepackage{amssymb}

\begin{document}

\title[An application of nonlinear supratransmission]{An application of nonlinear supratransmission to the propagation of binary signals in weakly damped, mechanical systems of coupled oscillators}

\author{J. E. Mac\'{\i}as-D\'{\i}az}
\address{Departamento de Matem\'{a}ticas y F\'{\i}sica, Universidad Aut\'{o}noma de Aguascalientes, Aguascalientes, Ags. 20100, Mexico}
\address{Department of Physics, University of New Orleans, New Orleans,  LA 70148}
\email{jemacias@correo.uaa.mx}

\author{A. Puri}
\address{Department of Physics, University of New Orleans, New Orleans,  LA 70148}
\email{apuri@uno.edu}


\date{\today}

\begin{abstract}
In the present article, we simulate the propagation of binary signals in semi-infinite, mechanical chains of coupled oscillators harmonically driven at the end, by making use of the recently discovered process of nonlinear supratransmission. Our numerical results --- which are based on a brand-new computational technique with energy-invariant properties --- show an efficient and reliable transmission of information.
\end{abstract}

\maketitle

The process of nonlinear supratransmission consists of a sudden increase in the amplitude of wave signals transmitted into a nonlinear chain by a harmonic disturbance at the end, irradiating at a frequency in the forbidden band gap. The phenomenon was first discovered in mechanical chains of oscillators described by coupled sine-Gordon and Klein-Gordon equations \cite{Geniet-Leon}, and it was quickly  studied in other nonlinear models \cite{Geniet-Leon2,Khomeriki,Leon-Spire,Khomeriki-Leon}. Several applications of nonlinear supratransmission have been realized so far \cite{Khomeriki-Leon2,Chevriaux}, and more physical applications have been given a mathematical foundation to its realization \cite{Khomeriki2,Khomeriki-Ruffo} have been suggested; however, the problem of transmitting coded information into nonlinear chain systems has not been attacked at all.

In this paper, we develop an application of nonlinear supratransmission to the propagation of binary signals in weakly damped semi-infinite mechanical chains of coupled oscillators, by modulating the amplitude of the driving signal at the end. The first section introduces the mathematical model under study and the energy expressions to be used. A numerical study of the idealized problem is carried out in the following section; here we provide a numerical analysis of the breather propagation, based on the method proposed in \cite{Macias-Supra}, which in turn is based on a numerical scheme to approximate radially symmetric solutions of modified Klein-Gordon equations \cite {Macias-Puri}. The next section presents a simulation of the propagation of a binary signal in a weakly damped system. Finally, we present a section of concluding remarks and propose complementary directions of research.

\section{Preliminaries}

\subsection{Mathematical model}

Throughout this paper, we assume that $\alpha$, $\beta$ and $\gamma$ are nonnegative real numbers, and that $c \gg 1$ (see \cite{Geniet-Leon}). Likewise, we consider a system $( u _n ) _{n = 1} ^\infty$ of oscillators satisfying the mixed-value problem studied in \cite{Macias-Supra}, namely,
\begin{equation}
\begin{array}{c}
\displaystyle {\frac {d ^2 u _n} {d t ^2} - \left( c ^2 +\alpha \frac {d} {d t} \right) \Delta ^2 _x u _n + \beta \frac {d u _n} {d t} + V ^\prime (u _n) = 0, }\\
		\begin{array}{rl}
        \begin{array}{l}
            {\rm subject\ to:} \qquad \\ \\ \\
        \end{array}
        \left\{
        \begin{array}{ll}
            u _n (0) = 0, & n \in \mathbb {Z} ^+, \\
            \displaystyle {\frac {d u _n} {d t} (0) = 0}, & n \in \mathbb {Z} ^+, \\
            u _0 (t) = \psi (t), & t \geq 0,
        \end{array}\right.
    \end{array}
\end{array}\label{Eqn:DiscreteMain}
\end{equation}
where $c$ is the coupling coefficient, and $\alpha$ and $\beta$ evidently play the roles of internal and external damping coefficients, respectively. Here, $\Delta ^2 _x u _n$ is used to denote the spatial second-difference $u _{n + 1} - 2 u _n + u _{n - 1}$ for every $n \in \mathbb {Z} ^+$, the boundary-driving function is given by $\psi (t) = A (t) \sin (\Omega t)$ for every $t \in (0 , + \infty)$, and $V (u _n) = 1 - \cos (u _n) - \gamma u _n$ where, due to the analogy with the Josephson model \cite {Zant}, $\gamma$ will be called here the \emph {normalized current}. Notice that the Hamiltonian of the $n$-th lattice site is given by 
$$
H _n = \frac {1} {2} \left[ \dot {u} _n ^2 + c ^2 (u _{n + 1} - u _n) ^2 \right] + V (u _n),
$$
for any differentiable function $V$. After including the potential energy from the coupling between the first two oscillators, the total energy of the system becomes
$$
E = \sum _{n = 1} ^\infty H _n + \frac {c ^2} {2} (u _1 - u _0) ^2.
$$

\subsection{Numerical schemes}

\begin{figure}[tcb]
\centerline{
\includegraphics[width=0.6\textwidth]{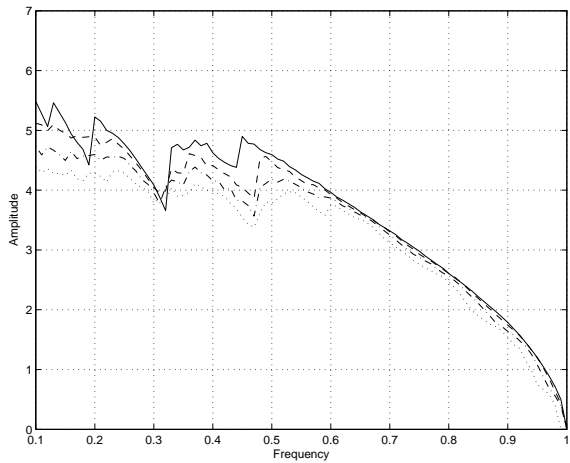}}
\caption{Bifurcation diagram of critical amplitude vs. driving frequency for system (\ref {Eqn:DiscreteMain}) for various values of $\gamma$: $0$ (solid), $0.1$ (dashed), $0.2$ (dash-dotted), $0.3$ (dotted). \label{Fig:Paper3Fig2}} %
\end{figure}

We consider a finite system of $N$ differential equations satisfying (\ref {Eqn:DiscreteMain}), and a regular partition $0 = t _0 < t _1 < \dots < t _M = T$ of the time interval $[0 , T]$ with time step equal to $\Delta t$. For each $k = 0 , 1 , \dots , M$, let us represent the approximate solution to our problem on the $n$-th lattice site at time $t _k$ by $u ^k _n$. If we convey that $\delta _t u _n ^k = u _n ^{k + 1} - u _n ^{k - 1}$, that $\delta ^2 _t u _n ^k = u _n ^{k + 1} - 2 u _n ^k + u _n ^{k - 1}$ and that $\delta ^2 _x u _n ^k = u _{n + 1} ^k - 2 u _n ^k + u _{n - 1} ^k$, the differential equations in our problem take then the discrete form

\begin{equation}
\displaystyle {\frac {\delta ^2 _t u _n ^k} {(\Delta t) ^2} - \left( c ^2 + \frac {\alpha} {2 \Delta t} \delta _t \right) \delta ^2 _x u _n ^k + \frac {\beta ^\prime} {2 \Delta t} \delta _t u _n ^k + \frac {V (u _n ^{k + 1}) - V (u _n ^{k - 1})} {u _n ^{k + 1} - u _n ^{k - 1}}} = 0,
\label{Eqn:DiscreteMainDiscr}
\end{equation}
where $\beta ^\prime$ includes both the effect of external damping and a simulation of an absorbing boundary slowly increasing in magnitude on the last $N - N _0$ oscillators. More concretely, we let $u _{N + 1} (t)$ be equal to zero at all time $t$, and let $\beta ^\prime$ be the sum of external damping and the function 
$$
\beta ^{\prime \prime} (n) = 0.5 \left[ 1 + \tanh \left( \displaystyle {\frac {2 n - N _0 + N} {6}} \right) \right],
$$
where usually $N _0 = 50$ and $N \geq 200$.

\begin{figure}[t]
\centerline{
\includegraphics[width=0.6\textwidth]{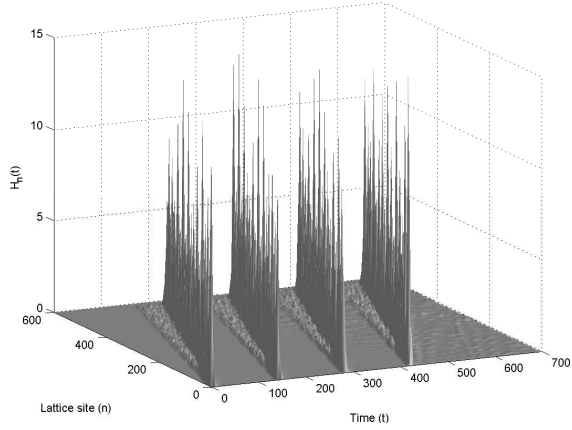}}
\caption{Local energies $H_n (t)$ of (\ref {Eqn:DiscreteMain}) vs. lattice site $n$ and time $t$, corresponding to the transmission of binary signal `$1111$'. \label{Fig:Paper3Fig4}} %
\end{figure}

Finite-difference scheme (\ref {Eqn:DiscreteMainDiscr}) (which is a modified version of the one proposed in \cite{Macias-Puri}) is consistent with our mixed-value problem and conditionally stable, having the inequality $\left(c \Delta t \right) ^2 < 1 + \left(\alpha + \beta ^\prime / 4\right) \Delta t$ as a necessary condition for stability when $\beta ^\prime$ is assumed constant \cite{Macias-Supra}. Moreover, if the energy of the system at the $k$-th time step is computed using the expression 
\begin{eqnarray*}
E _k & = & \frac {1} {2} \sum _{n = 1} ^ M \left( \frac {u _n ^{k + 1} - u _n ^k} {\Delta t} \right) ^2 + \frac {c ^2} {2} \sum _{n = 1} ^M (u _{n + 1} ^{k + 1} - u _n ^{k + 1}) (u _{n + 1} ^k - u _n ^k) \\
 & & \qquad +\sum _{n = 1} ^M \frac {V (u _n ^{k + 1}) + V (u _n ^k)} {2} + \frac {c ^2} {2} (u _1 ^{k + 1} - u _0 ^{k + 1}) (u _1 ^k - u _0 ^k),
\end{eqnarray*}
then the discrete rate of change of energy turns out to be a consistent approximation of order $\mathcal {O} (\Delta t) ^2$ for the corresponding instantaneous rate of change.

\section{Numerical study\label{Sec3}}

\subsection{Bifurcation analysis}

\begin{figure}[tcb]
\centerline{
\includegraphics[width=0.6\textwidth]{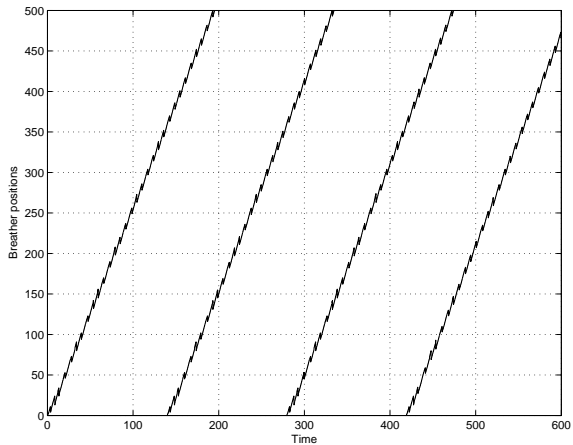}}
\caption{Time-dependent graphs of the position of the breathers generated in (\ref {Eqn:DiscreteMain}) by the binary signal `$1111$'. \label{Fig:Paper3Fig5}} %
\end{figure}

The existence of a bifurcation threshold of the energy administered into a semi-infinite chain of damped coupled oscillators described by (\ref {Eqn:DiscreteMain}) has been established and numerically predicted in \cite{Macias-Supra} for a potential $V (u) = 1 - \cos u$. Numerical experiments on undamped mechanical chains with nonzero normalized bias currents have shown that the process of nonlinear supratransmission is likewise present in these models. In fact, Fig. \ref{Fig:Paper3Fig2} provides bifurcation diagrams of driving amplitude at which supratransmission first occurs vs. driving frequency $\Omega$ for a system of $200$ undamped oscillators with coupling coefficient equal to $4$ and several values of $\gamma$, over a time interval $[0 , T (\Omega)]$ where $T (\Omega)$ is equal to $200$ for all frequencies except for those satisfying $\Omega > 0.95$, in which case $T (\Omega)$ had to be increased up to $500$.

\subsection{Moving breather solutions\label{SubSec32}}

For the sake of simplification, let $\gamma = 0$ and consider a discrete system described by (\ref{Eqn:DiscreteMain}), harmonically driven at the boundary by a frequency $\Omega$. A binary bit $b$ will be transmitted into the medium during a fixed and sufficiently long period of signal generation $P$ equal to an integer multiple of the driving period, by defining
$$
A (t) = \frac {8} {5} b C A_s \left(e ^{- \Omega t / 4.5} - e ^{- \Omega t / 0.45} \right)
$$
for every $t \in [0 , P]$, where $A_s$ represents the critical amplitude at which nonlinear supratransmission starts and $C > 0$ is an \emph {adjusting constant}. 

\begin{figure}[t]
\centerline{
\includegraphics[width=0.6\textwidth]{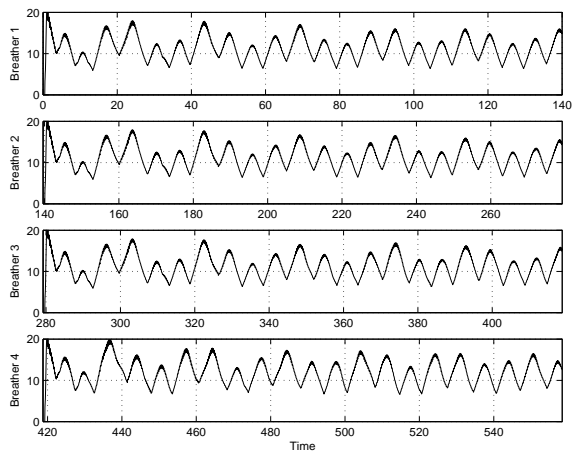}}
\caption{Time-dependent graphs of the maximum energy of the breathers generated in (\ref {Eqn:DiscreteMain}) by the binary signal `$1111$'. \label{Fig:Paper3Fig6}} %
\end{figure}

Fig. \ref{Fig:Paper3Fig4} shows the time evolution of the local energies $H _n$ corresponding to solutions of the undamped problem (\ref {Eqn:DiscreteMain}) with coupling coefficient equal to $4$, driving frequency $0.9$ (in which case $A _s$ is approximately equal to $1.79$) and $C = A _s$. The period of signal generation is equal to $20$ driving periods, and the binary code transmitted at the boundary is `$1111$'. The results show that a single moving breather is generated per period, and that the phase velocity $v _p$ through the medium is approximately constant. To verify this claim, we include in Fig. \ref{Fig:Paper3Fig5} the graphs of the positions of the four breathers generated by the binary signal vs. time. The fact that the breathers are transmitted at a constant velocity is now obvious, the common phase velocity being approximately $2.549$.

In order to determine the strength of the emitted signals at sites located far away from the source, it is important to determine the time behavior of the maximum local energy attained by a discrete breather. Fig. \ref{Fig:Paper3Fig6} presents this behavior for the breather solutions obtained under the conditions above. Notice that the maximum local energy is well above a cutoff limit of $2$.

\section{Application}

The discrete medium described by (\ref {Eqn:DiscreteMain}) with $\alpha = \beta = \gamma = 1 \times 10 ^{- 3}$, and the rest of the parameters as in the previous section, will be our object of study in this section. Local energy-based reception devices will be placed on the $100$-th and $300$-th lattice sites, and the signal to be transmitted through the chain system is `$10111001011011101001$'. Our system will consist of $600$ sites, and a time step of $0.05$ will be employed.

\begin{figure*}[t]
\centerline{
\begin{tabular}{cc}
\includegraphics[width=0.45\textwidth]{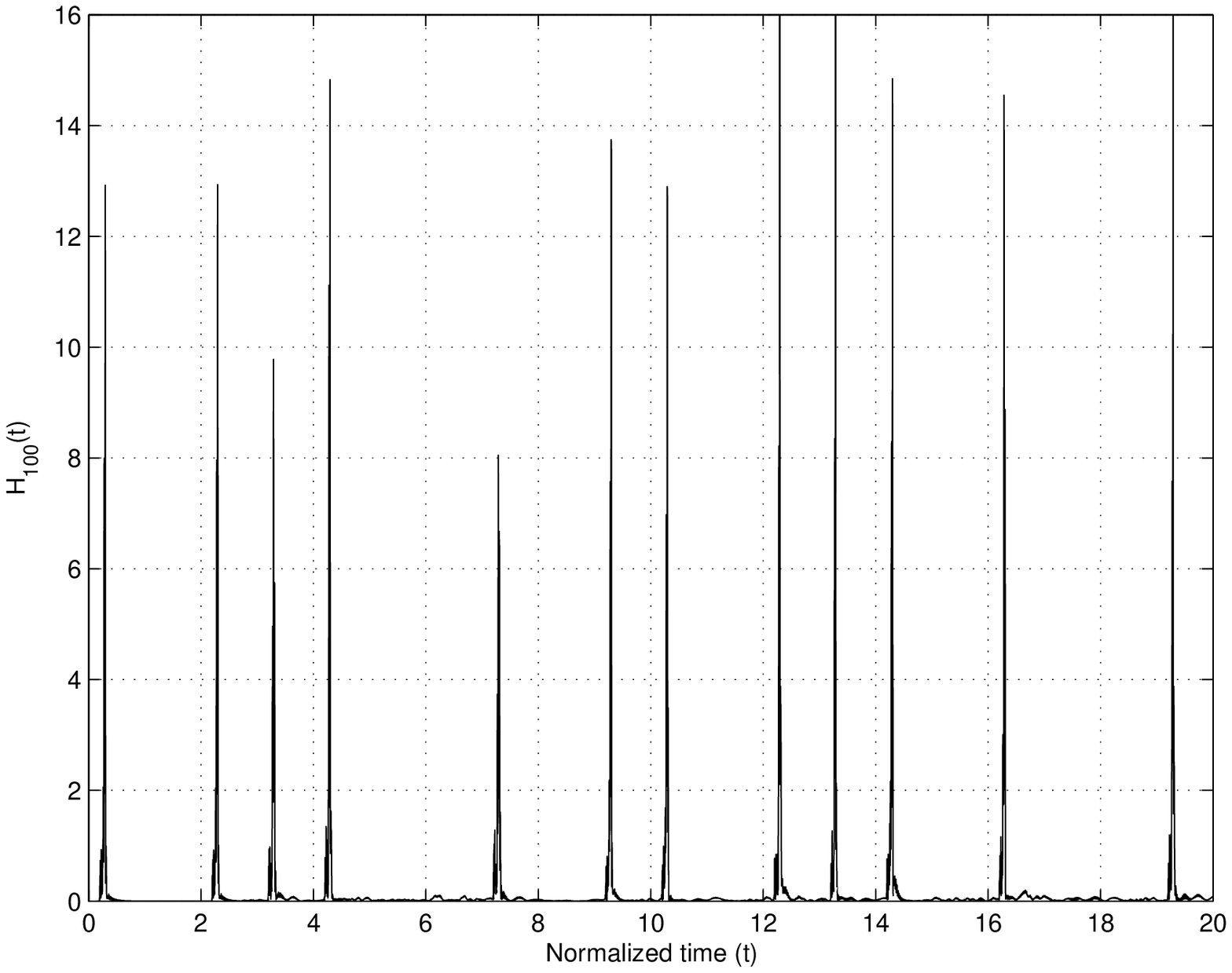}&
\includegraphics[width=0.45\textwidth]{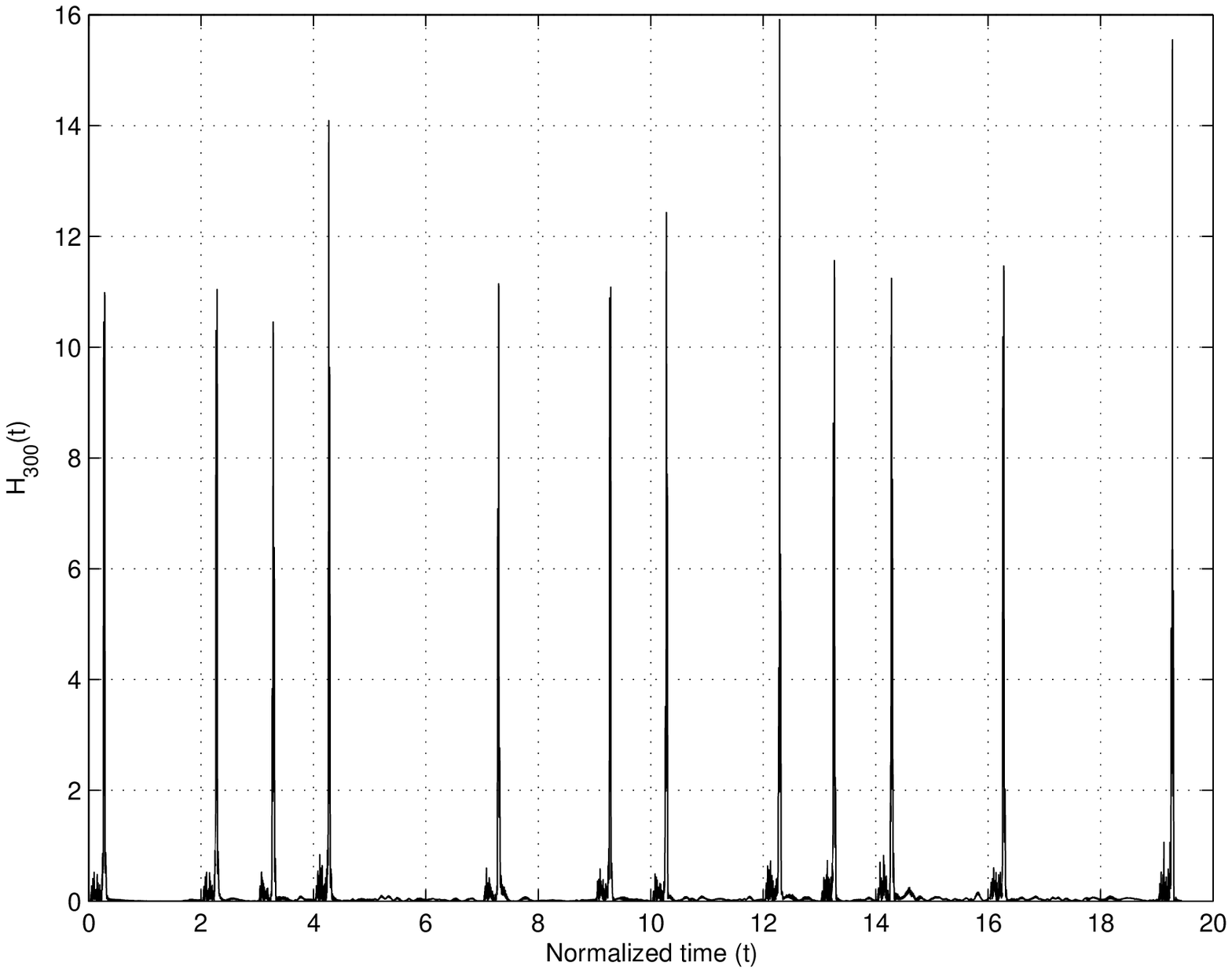}
\end{tabular}}
\caption{Local energy of the $100$-th (left) and $300$-th (right) sites in (\ref {Eqn:DiscreteMain}) vs. normalized time, as a response to the transmission of the binary signal `$10111001011011101001$'. \label{Fig:FirstSignal}} %
\end{figure*}

Assuming the general convention that site $n _0 \geq 100$ will start signal reception at time $t = (n _0 - 100) / v _p$, Fig. \ref{Fig:FirstSignal} shows the evolution of the local energy in the $100$-th and $300$-th sites in terms of the time normalized with respect to the period of signal generation. In either case, it is clear that the transmission of a bit equal to $1$ in the $n$-th period is completely characterized in the graph by peak(s) of height greater than the cutoff limit $2$ on the interval $[n - 1 , n]$.

\section{Conclusions and perspectives}

In this letter, we have proposed a simple mathematical model to transmit binary information in discrete, semi-infinite chains of couped oscillators using the process of nonlinear supratransmission. In the absence of dispersive and dissipative effects, our model (which is based on the modulation of amplitudes of source signals with constant frequency) has shown to be highly reliable for sufficiently long periods of single-bit generation, independently of the distance between the source of transmission and the point of reception. 

When weak damping is present the general picture does not change much. Stronger damping, however, manifests itself through a substantial decrease in the amplitude of the maximum local energy of the moving breathers with respect to the lattice site. In a forthcoming work, we will examine the possibility to overcome this problem via the concatenation of chain systems, where the driving at the beginning of each lattice will irradiate with an amplitude equal to a value just below its critical point multiplied by the amplitude of the last site in the previous lattice.

Finally, we wish to point out that the problem of determining whether it is possible to design a propagation system of binary signals in Josephson junction arrays still remains an open topic of research. Moreover, in view of the recently discovered phenomenon of nonlinear infratransmission (or lower-transmission, as named by the authors \cite{Khomeriki}), the problem of finding more efficient pathways to achieve signal transmission in other models is still an open question of general interest.

\subsubsection*{Acknowledgments}

One of us (J. E. M. D.) wishes to express his most sincere gratitude to Dr. \'{A}lvarez Rodr\'{\i}guez, dean of the Centro de Ciencias B\'{a}sicas of the Universidad Aut\'{o}noma de Aguascalientes, and to Dr. Avelar Gonz\'{a}lez, head of the Direcci\'{o}n General de Investigaci\'{o}n y Posgrado of the same university, for providing him with the physical means to produce this article. He also wishes to acknowledge enlightening conversations with Prof. F. Rizo D\'{\i}az in the Departamento de Sistemas Electr\'{o}nicos. The present work represents a set of partial results under project PIM07-2 at this university.

\end{document}